\documentclass{iaus}
\usepackage{graphicx}

\title[]{3D model of magnetic fields evolution in dwarf irregular galaxies}

\author[Siejkowski et al.]
  {Hubert Siejkowski$^1$,
   Marian Soida$^1$,
   Katarzyna Otmianowska-Mazur$^1$,
   Micha\l{} Hanasz$^2$ \and
   Dominik J. Bomans$^3$}

\affiliation{
  $^1$Astronomical Observatory, Jagiellonian University, ul. Orla 171, 30-244 Krak\'{o}w, Poland\\
  email: {\tt h.siejkowski@oa.uj.edu.pl} \\[\affilskip]
  $^2$Torun Centre for Astronomy, Nicolaus Copernicus University, 87-148 Toru\'{n}/Piwnice, Poland\\[\affilskip]
  $^3$Astronomical Institute of Ruhr-University Bochum, Univeristatsstr. 150/NA7, D-44780 Bochum, Germany}

\pubyear{2010}
\volume{274}
\jname{Advances in Plasma Astrophysics}
\editors{A. Bonanno, E. de Gouveia Dal Pino \& A. Kosovichev, eds.}

\begin{document}

\maketitle

\begin{abstract} 

Radio observations show that magnetic fields are present in dwarf irregular galaxies (dIrr) and its
strength is comparable to that found in spiral galaxies. Slow rotation, weak shear and shallow
gravitational potential are the main features of a typical dIrr galaxy.  These conditions of the
interstellar medium in a dIrr galaxy seem to unfavourable for amplification of the magnetic field
through the dynamo process.  Cosmic-ray driven dynamo is one of the galactic dynamo model, which has
been successfully tested in case of the spiral galaxies. We investigate this dynamo model in the ISM
of a dIrr galaxy. We study its efficiency under the influence of slow rotation, weak shear and
shallow gravitational potential. Additionally, the exploding supernovae are parametrised by the
frequency of star formation and its modulation, to reproduce bursts and quiescent phases. We found
that even slow galactic rotation with a low shearing rate amplifies the magnetic field, and that
rapid rotation with a low value of the shear enhances the efficiency of the dynamo. Our simulations
have shown that a high amount of magnetic energy leaves the simulation box becoming an efficient
source of intergalactic magnetic fields.
  
\end{abstract} 
\section{Introduction} 
\label{sec:Introduction}

Dwarf irregular galaxies have relatively simple structure. They are smaller, less massive and have
lower luminosity than spirals and ellipticals. Their rotation speed is very low and the rotation
curve could be very complex (e.g. NGC 4449). The structure of a galaxy is often disturbed by a
strong burst of star formation. Weak gravitational potential and slow rotation cause that supernovae
explosions can substantially influence the gas distribution and global velocity pattern. Energy
injected by a starbursting events is enough to drive a gas outflow from a dwarf galaxy. Together
with gas also metals and magnetic fields are transported.
  
Radio observations (Chy\.zy et al. 2000, 2003; Kepley et al. 2010; Klein et al. 1991, 1992) show that
dIrr galaxies can have relatively strong magnetic fields. The typical total magnetic field strength
is 5--15~$\mu$G with a uniform component about 5~$\mu$G. The observed magnetic fields suggest that a
dynamo process should operate in these galaxies. We investigate the cosmic-ray driven dynamo model
in the environment of a typical dwarf irregular galaxy.

\section{Model description and initial setup} 
\label{sec:Model description}

The CR-driven dynamo model consists of the following elements based on Hanasz et al. (2006) and
references therein. We assume:
\begin{itemize}
\item 
the cosmic ray component described by the~diffusion-advection transport equation and we adopt
anisotropic diffusion;
\item 
localized sources of CR, i.e. random explosions of supernovae in the disk volume. The cosmic ray
input of individual SN remnant is $10\%$ of the canonical kinetic energy output
($10^{51}~\textrm{erg}$);
\item 
resistivity of the ISM to enable the dissipation of the small-scale magnetic fields (see
Hanasz \& Lesch 2003). In the model, we apply the uniform resistivity and neglect the
Ohmic heating of gas by the resistive dissipation of magnetic fields;
\item 
shearing boundary conditions, tidal and Coriolis forces;
\item 
realistic vertical disk gravity following the model by Ferri\`{e}re (1998) with rescaled
disk and halo masses by one order of magnitude.
\end{itemize}

The 3D cartesian domain size is $0.5~\textrm{kpc}\times1~\textrm{kpc}\times8~\textrm{kpc}$ in
$x,y,z$ coordinates corresponding to the radial, azimuthal, and vertical directions, respectively,
with a grid size $(20~\textrm{pc})^3$. The boundary conditions are sheared-periodic in $x$, periodic
in $y$, and outflow in $z$ direction. The positions of SNe are chosen randomly with a~uniform
distribution in the $xy$ plane and a~Gaussian distribution in the vertical direction. In addition,
the SNe activity is modulated during the simulation.  The applied value of the perpendicular CR
diffusion coefficient is $K_\perp = 10^3~\textrm{pc}^2~\textrm{Myr}^{-1}$ and the parallel one is
$K_\parallel = 10^{4}~\textrm{pc}^2~\textrm{Myr}^{-1}$ (see Hanasz et al. 2009). The initial state
of the system represents the magnetohydrostatic equilibrium with the horizontal, purely azimuthal
magnetic field with $p_{\it mag}/p_{\it gas} = 10^{-4}$.

\section{Results} 
\label{sec:Results}

\subsection{Rotation and shear} 
\label{sub:Rotation and shear}

\begin{figure*}[t]
\centering
\includegraphics[width=0.46\textwidth,keepaspectratio=true]{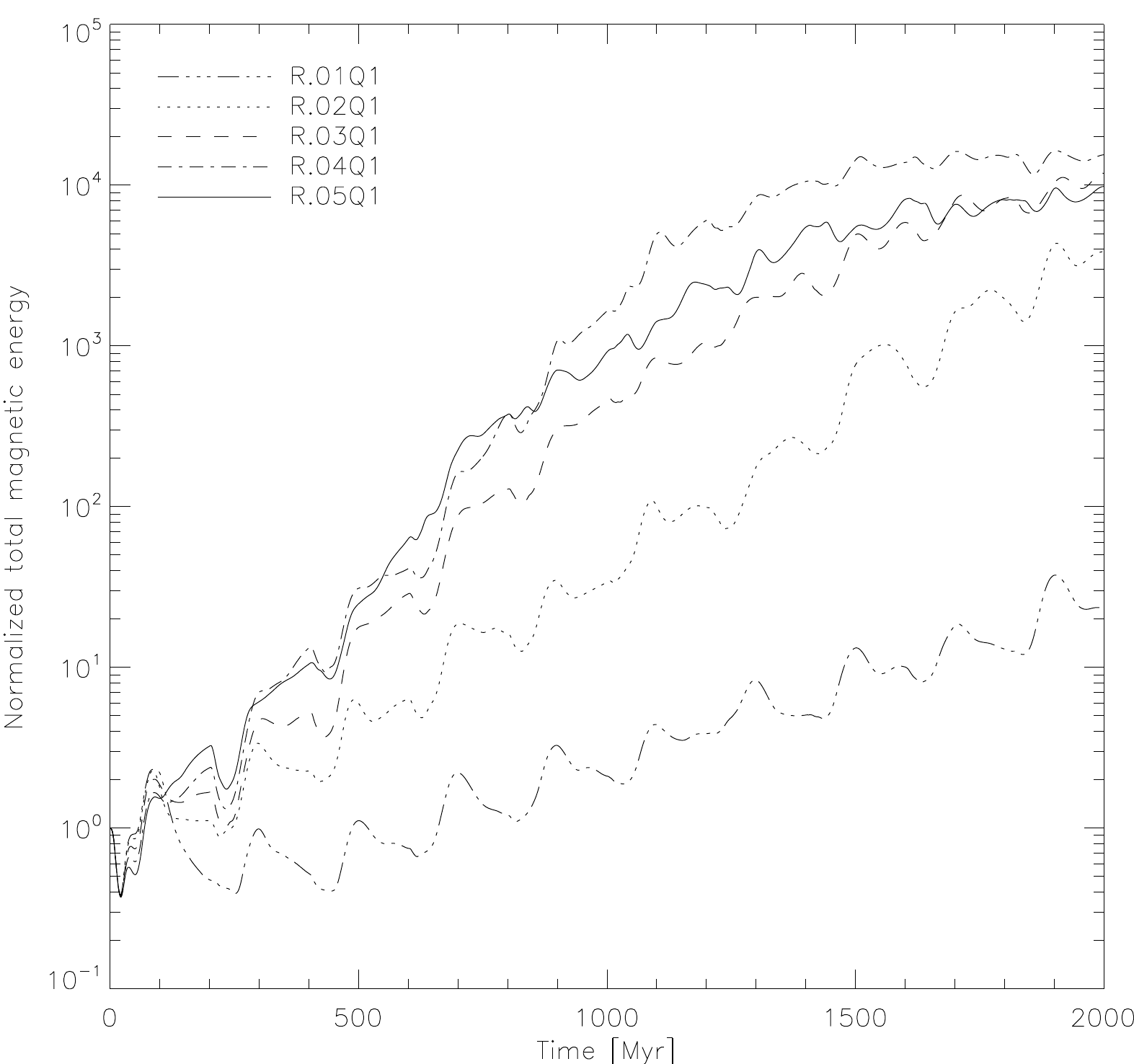}
\hspace{0.2cm}
\includegraphics[width=0.46\textwidth,keepaspectratio=true]{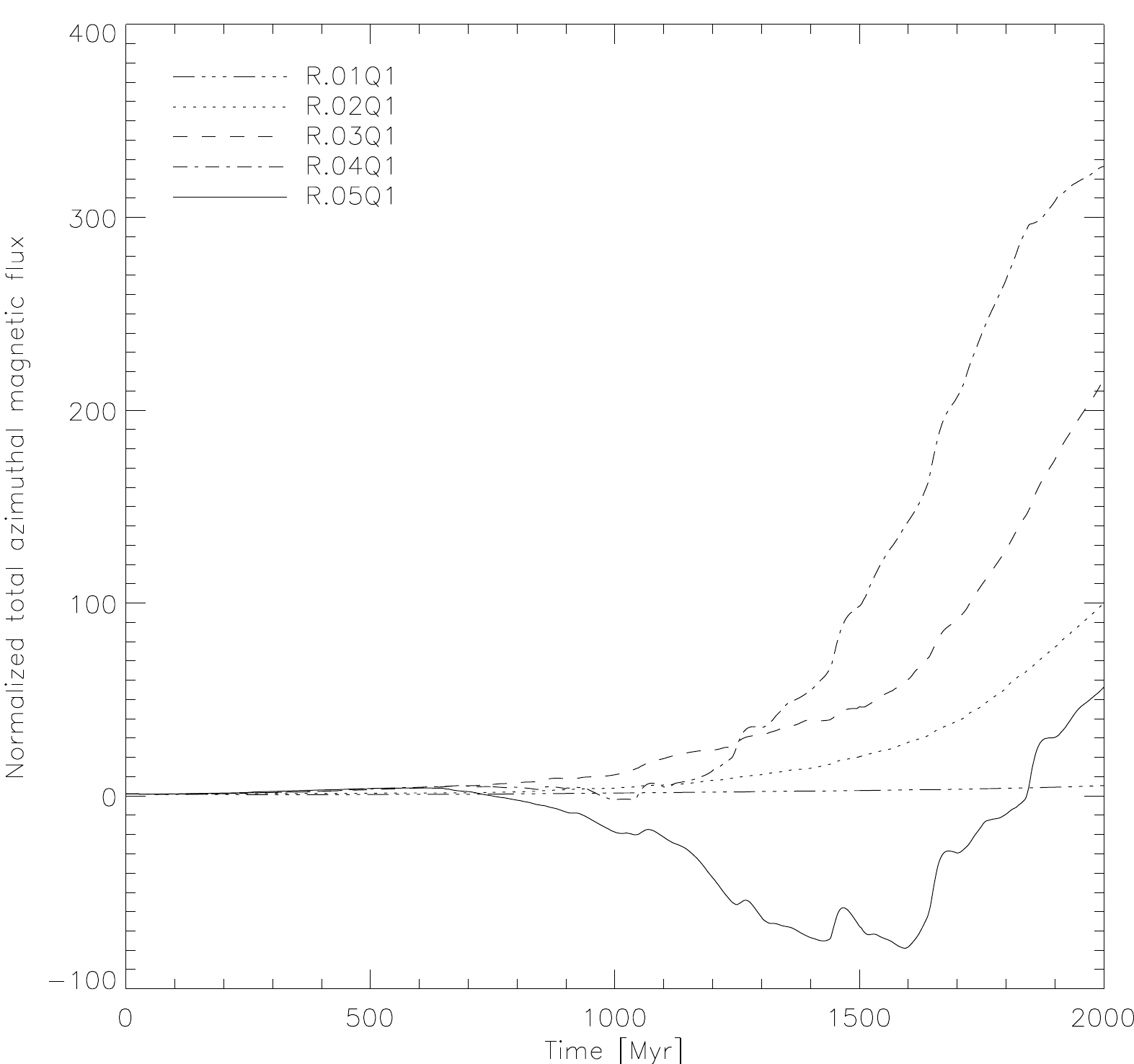}
\caption{Evolution of the total magnetic energy $E_B$ (left panel) and the total azimuthal flux $B_\phi$
(right) for models with different rotation. Both quantities are normalized to the initial value.}
\label{fig:rotation}
\end{figure*}

We studied dependence of the magnetic field amplification on the parameters describing the rotation
curve, namely, the shearing rate $q$ and the angular velocity $\Omega$. The evolution in the total
magnetic field energy $E_B$ and total azimuthal flux $B_\phi$ for different values of $\Omega$ is
shown in Fig.~\ref{fig:rotation}, left and right panel, respectively. Models with higher angular
velocities, starting from 0.03~Myr$^{-1}$, initially exhibit exponential growth of the magnetic
field energy $E_B$ till 1\,200~Myr followed by a saturation.  The saturation values of the magnetic
energy for these three models are similar and $E_B$ exceeds the value $10^4$ in the normalized
units. The magnetic energy in the models R.01Q1\footnote{Letter R stands for angular velocity
(rotation) given in Myr$^{-1}$, Q shearing rate ($q=-d\ln \Omega/ d\ln R$) and SF for star formation
given in kpc$^{-2}$~Myr$^{-1}$.} (slow rotation, moderate shear) and R.02Q1 (slow rotation, moderate
shear) grows exponentially during the whole simulation and does not reach the saturation level. The
total azimuthal magnetic flux evolution (Fig.~\ref{fig:rotation}, right) shows that a higher angular
velocity leads to a higher amplification. There is no amplification in model R.01Q1.

\subsection{Star formation} 
\label{sub:Star formation}

\begin{figure*}[t]
\centering
\includegraphics[width=0.46\textwidth,keepaspectratio=true]{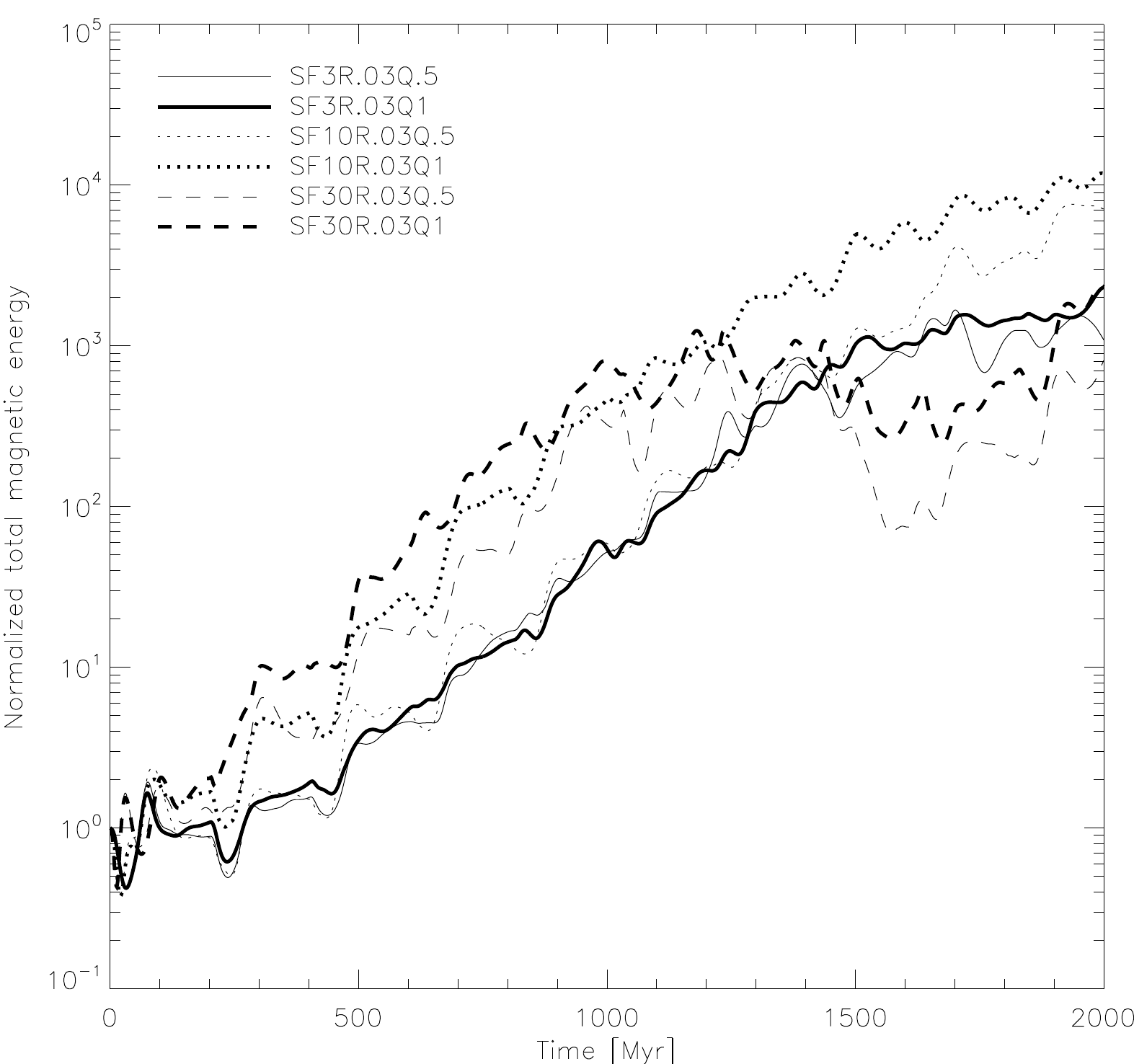}
\hspace{0.2cm}
\includegraphics[width=0.46\textwidth,keepaspectratio=true]{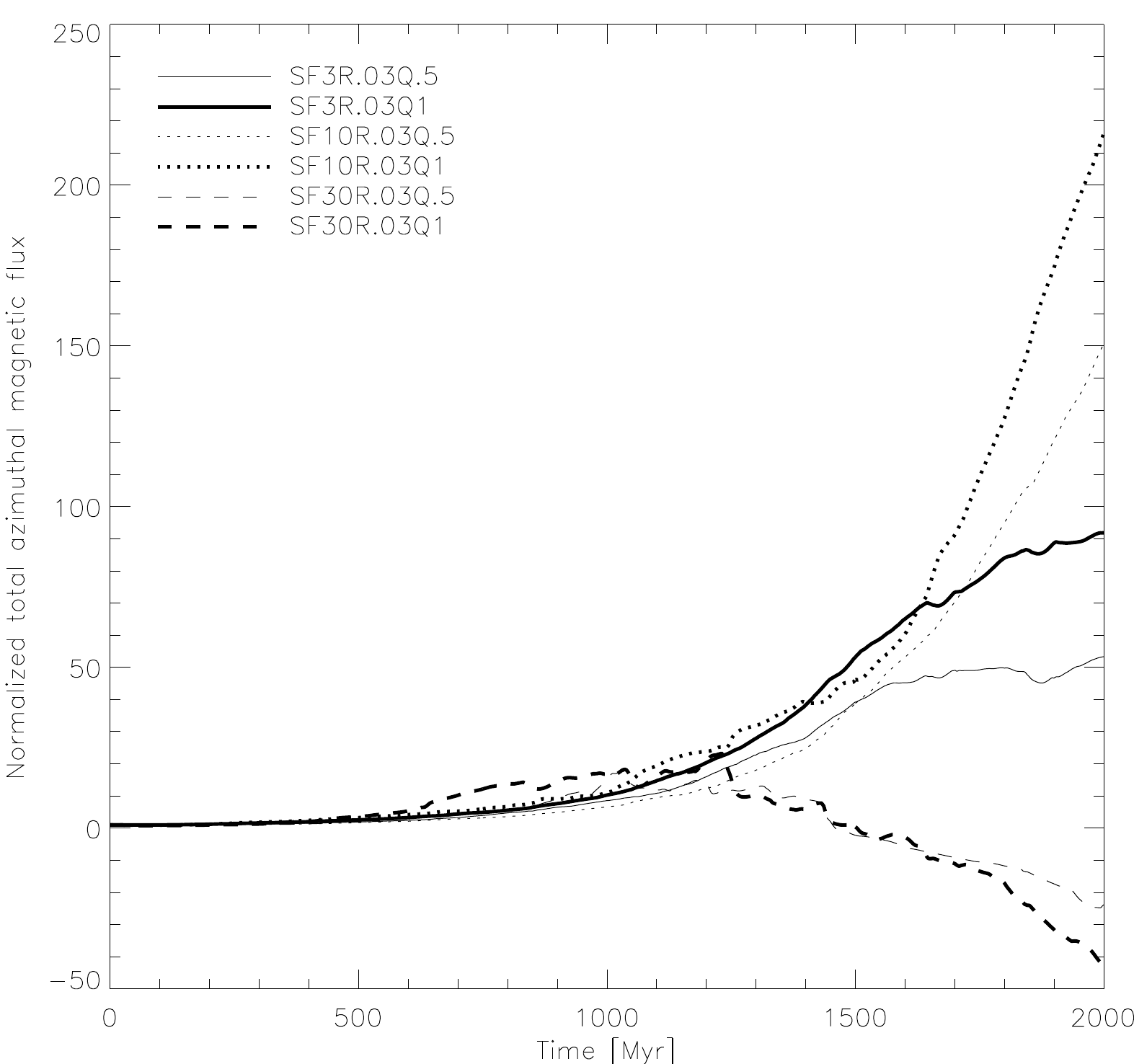}
\caption{Evolution of the total magnetic energy $E_B$ (left panel) and the total azimuthal flux
$B_\phi$ (right) for models with different supernova explosion frequency and shearing rate. Both
quantities are normalized to the initial value.} \label{fig:frequency}
\end{figure*}

We checked how the frequency and modulation of SNe influence the amplification of magnetic fields.
The evolution in total magnetic field energy and total azimuthal flux for different supernova
explosion frequencies are shown in Fig. \ref{fig:frequency}. The total magnetic energy evolution for
all models is similar, but the differences are apparent in the evolution of azimuthal flux.  The
most efficient amplification of $B_\phi$ appears for SF10R.03Q.5 (medium SFR, moderate rotation, low
shear) and SF10R.03Q1 (moderate shear), and for other models the process is less efficient.  In
addition, for models SF30R.03Q.5 (high SFR, low shear) and SF30R.03Q1 (high SFR, moderate shear), we
observe a turnover in magnetic field direction. The results suggest that the dynamo requires higher
frequencies of supernova explosions to create more regular fields, although, if the explosions occur
too frequently because of a strong wind transporting magnetic field out of the disk.

\subsection{Outflow of magnetic field} 
\label{sub:Outflow of magnetic field}

To measure the total production rate of the magnetic field energy during the simulation time, we
calculated the outflowing $E_B^{\it out}$ through the $xy$ top and bottom domain boundaries. To
estimate the magnetic energy loss, we computed the vertical component of the Poynting vector.  Its
value is computed in every cell belonging to the top and bottom boundary planes and then integrated
over the entire area and time.  For models with a~low dynamo efficiency most of the initial magnetic
field energy is transported out of the simulation box. In some cases (i.e., all models except R.01Q0
and R.05Q0 with zero shear), we find that the energy loss $E_B^{\it out}$ is comparable to the energy
remaining inside the domain $\bar{E}_B^{\it end}$. In these models, the ratio $E_B^{\it
out}/\bar{E}_B^{\it end}$ varies from 0.03 to 0.96 and is highly dependent on the supernova
explosion frequency.  The results show that the outflowing magnetic energy is substantial (see
Siejkowski et al. 2010) suggesting, that irregular galaxies can be efficient sources of
intergalactic magnetic fields.


\section{Conclusions} 
\label{sec:Conclusion}

We have described the evolution of the magnetic fields in irregular galaxies in terms of a
cosmic-ray driven dynamo (Siejkowski et al. 2010). The amplification of magnetic fields have been
studied under different conditions characterized by the rotation curve (the angular velocity and the
shear) and the supernovae activity (its frequency and modulation) typical for irregular galaxies. We
have found that:

\begin{itemize}
\item in the presence of slow rotation and weak shear in irregular galaxies, the amplification
of the total magnetic field energy is still possible;
\item shear is necessary for efficient action of CR-driven dynamo, but the amplification itself
depends weakly on the shearing rate;
\item higher angular velocity enables a higher efficiency in the CR-driven dynamo process;
\item the efficiency of the dynamo process increases with SNe activity, but excessive SNe
activity reduces the amplification;
\item for high SNe activity and rapid rotation, the azimuthal flux reverses its direction;
\item the outflow of magnetic field from the disk is high, suggesting that dIrr galaxies may
magnetize the intergalactic medium as predicted by Kronberg et al.  (1999) and Bertone et al.
(2006).
\end{itemize}

The performed simulations indicate that the CR-driven dynamo can explain the observed magnetic
fields in dwarf irregular galaxies. In future work we plan to determine the influence of other ISM
parameters and perform global simulations of these galaxies.

\acknowledgements{
This work was supported by Polish Ministry of Science and Higher Education through grants:
92/N-ASTROSIM/2008/0 and 3033/B/H03/2008/35, and by the DFG Research Group FOR1254. Presented
computations have been performed on the GALERA supercomputer in TASK ACC in Gda\'nsk.}

\end{document}